\magnification=1200

\def\cc{C/C_{lim}}

\def\vavg{\langle V/V_{max}\rangle} 
 
\def\om{\Omega} 
\def\Om{\Omega}
\def\vm{V/V_{max}}

\def\lnc{\log N-\log C} 

\leftline{\bf Cosmological parametrization of gamma ray burst models} 
\leftline{Eric V. Linder} 
\bigskip 
\leftline{Imperial College, Prince Consort Road, London SW7 2BZ, 
England; el@ic.ac.uk} 
\bigskip\bigskip 

\noindent{\bf Abstract.} 
Using three parametrizations of the gamma ray burst count data comparison 
is made to cosmological source models.  While simple models can fit 
$\vavg$ and faint end slope constraints, the addition of a logarithmic 
count range variable describing the curvature of the counts shows that 
models with no evolution or evolution power law in redshift with index 
less than 10 fail to satisfy simultaneously all three 
descriptors of the burst data.  The cosmological source density that 
would be required for a fit is illustrated. 
\bigskip\medskip 

\leftline{\bf 1. Introduction} 
\medskip\noindent 
Gamma ray bursts present an astrophysical puzzle as to their mechanism, 
source, and even locality.  Despite knowledge of their existence for 
almost 25 years and over 1500 detected bursts, we still do not know 
if they belong to the realm of the solar system, galaxy, or high 
redshift cosmology, a state that would be unthinkable for other initially 
mysterious sources such as quasars.  In the absence of a sufficiently 
robust clue from their time structure, energy spectrum, or a non gamma 
ray counterpart we turn to their spatial distribution.  Their angular 
distribution on the sky is isotropic to 1\% and no distance parametrization 
is known for them.  We therefore have recourse only to the number-flux 
relation, $\log N-\log S$, although for gamma ray bursts the peak count 
rate in the detector, C, is used instead of the energy flux $S$. 

Section 2 discusses the count distribution from the point of view 
of properties of the detector, including threshold bias, and of 
the cosmology, e.g.~geometry.  Section 3 introduces three 
parametrizations of the distribution for comparison of theoretical 
source models with the data.  The 
main, new results of this paper are illustrated by Figure 1 showing the 
lack of agreement between the data and those cosmological models 
with no evolution or evolution power law in redshift and by 
Figure 2 showing what form of cosmological source distribution would be 
required to fit the data. 
\bigskip 
\leftline{\bf 2. Count distribution} 
\medskip\noindent 
In analogy with other wavelength regimes (e.g.~radio) of astrophysics, 
one considers the data set composed of values of 
the peak counts $C$ and the count detection threshold $C_{lim}$.  One 
wants to find the number distribution $N(C)$ in the hopes of this 
giving a clue to the source model.  (For gamma ray bursts the count 
rate $C$ is used instead of the energy flux $S$). 
An advantage often accrues to using a 
relative variable $C/C_{lim}$, in that many instrumental biases cancel 
out in such a ratio (see, for example, Schmidt, Higdon, and Hueter 1988).  
From this basic relative variable one can form the ``volume fraction'' 
variable $\vm=(\cc)^{-3/2}$. 
However the effect of variation of the threshold value on recovering 
the entire distribution $N(C)$ is problematic.  Another difficulty is 
the nontrivial 
physical, i.e. spatial, interpretation of variables like this. 

\bigskip 
\leftline{\it 2.1. Source and detector variation} 
\medskip\noindent 
First we consider to what 
extent $v=\vm$ reflects the true source space distribution rather than 
effects arising from variations in source or detector properties.  In 
\S2.2 difficulties from the spatial geometry itself are discussed. 

Define the relation between detected counts $C$, source count 
luminosity $L$, 
and ``distance'' $r$ according to the usual inverse square law, 
$$r\equiv (L/4\pi C)^{1/2}.\eqno(1)$$ 
When $C$ is replaced by $C_{lim}$ this likewise defines the maximum 
``distance'' $D$ at which a source of (count) luminosity $L$ could be seen 
by a detector with count threshold $C_{lim}$.  Only when the source 
and detector properties, i.e. $L$ and $C_{lim}$, are known can $v=(r/D)^3$ 
be interpreted in terms of a true distance or spatial measure.  
A range of $L$ and $C_{lim}$ around some fiducial values will 
cloud our understanding of the true source 
radial distribution.  [Note that other variables could have a similar 
effect on interpreting $r$ as a distance.  For example, a distribution of 
source beaming properties, rather than isotropic emission, would affect 
the $4\pi$ factor in (1); energy spectrum variations enter the 
conversion of $L$ from an energy luminosity to a count rate; detector 
efficiency or sky coverage fluctuations alter the effective maximum 
volume, etc.] 

The distribution function $p(v)$ is the fraction of detectable sources 
lying in a (pseudo) volume interval: 
$$p(v)dv={\cal N}dV/N_{tot},\eqno(2)$$ 
where ${\cal N}$ is the source number density and $N_{tot}$ is the total 
number of detectable sources, normalizing the distribution.  Given 
dispersion in a property intrinsic to the source, say $L$, and one 
intrinsic to the detector, say $C_{lim}$, how is the distribution function 
affected relative to the standard candle, standard camera case?  In other 
words, how are the distributions that include the property 
dispersions, denoted $[p(v;L)]_C$, $[p(v;C_{lim})]_L$, and $[p(v)]_{L,C}$, 
related to the fiducial case $p(v;L,C_{lim})$?  The subscript notation 
indicates a weighted average with respect to that quantity, i.e. an 
integration over its variation. 

The procedure is most familiar in the case of a luminosity function 
$\Phi(L)$, 
with ${\cal N}=n(V)\Phi(L)dL$ and $n$ the spatial number density.  
One cannot simply write the integration 
over the dispersion as 
$$[p(v;C_{lim})]_L=\int dL\,\Phi(L)p(v;C_{lim},L),\eqno(3)$$ 
because of the dependence in the normalizing factor $N_{tot}$.  There is 
uniform weighting not in terms of the source fraction intrinsically 
existing at luminosity $L$, i.e. $\Phi(L)$, but in the fraction 
{\it detectable}, as the definition of $N_{tot}$ states.  That is, 
$$\eqalign{[p(v;C_{lim})]_L&=\int dL\,\Phi(L)n(V)(dV/dv)\Big/\int dL\, 
\Phi(L)\int dV\,n(V)\cr &=\int dL\,\Phi(L)N_{tot}(L,C_{lim}) p(v;C_{lim}, 
L)\Big/\int dL\,\Phi(L)\int dV\,n(V)\cr &=\int dL\,\psi(L,C_{lim}) 
p(v;C_{lim},L).\cr}\eqno(4)$$ 
Two alternative views of this procedure are equivalent: it is either a 
number weighting [by $\Phi(L)\,N_{tot}(L)/N_{tot}$] 
of the variable $p(v;C_{lim},L)=N_{tot}^{-1}(L)n(V)
(dV/dv)$, as in the last line, or a uniform weighting [by $\Phi(L)$] 
of the fundamental 
variable $N_{tot}^{-1}n(V)(dV/dv)$, as in the first line. 

The case of dispersion in the threshold $C_{lim}$ is completely 
analogous since again this variable enters in the normalization factor. 
Both dispersions together simply involve 
a double integration over the variables.  In this case $\Phi(L)$ is 
joined by the analogous $f(C_{lim})$ and the weighting function $\psi$ 
is generalized to 
$$\chi(C_{lim},L)=f(C_{lim})\Phi(L)N_{tot}(L,C_{lim})\Big/\int 
dC_{lim}f(C_{lim})\int dL\,\Phi(L) \int dV\,n(V).\eqno(5)$$ 
The effect is to introduce a coarse grained averaging or diffusivity over 
the distribution of $v$, smearing out sharp features.  Even worse, it can 
act as a redistribution mechanism, shifting high count rates (low $v$), 
for example, to low.  This is very similar to the case of amplification 
bias from gravitational lensing in flux limited surveys (cf.~Schneider 
1987).  The observed faint end slope of $-1$ in $N(C)$ 
could be obtained from a 
steeper true source distribution if there existed a broad (flat $f$) 
threshold variation.  Petrosian (1993; also see Caditz and Petrosian 
1993) deconvolves the threshold variations from the 
count data using a nonparametric maximum likelihood statistical method 
to derive a true faint end slope for $N(C)$ in the 
range $-1.6$ to $-1.8$.  The distribution is still reasonably 
fit by a broken power law, with a bright end slope of $-2.5$.  See 
\S3 for further discussion. 

\bigskip 
\leftline{\it 2.2. $\vm$ properties}
\medskip\noindent 
Because of its historical importance in the gamma ray burst debate we 
briefly consider the volume fraction variable $v\equiv\vm$, concentrating 
on its implications for the geometry 
of the source distribution, particularly cosmologically.  
The comoving volume element is 
$$dV=(1+z)^3r_a^2dr_{pr}d\omega,\eqno(6)$$ 
where $z$ is the redshift, 
$r_{pr}$ the proper distance, $r_a$ the angular diameter distance, 
and $d\omega$ the solid angle element.  Since observations indicate 
that bursts are distributed isotropically, $d\omega$ will simply 
contribute a factor $4\pi$. 

As known from gravitational lensing, when dealing with a line of sight 
toward a source one 
must use an angular diameter distance that incorporates the fact that 
this direction is clear 
of density inhomogeneities that might appreciably (i.e. detectably) 
lens and alter the flux. 
These are known as clumpy universe distances and are parametrized by 
the ratio of smoothly distributed matter density to the total (Dyer and 
Roeder 1973, Ehlers and Schneider 1986).  When this ratio is unity the 
angular distances agree with the standard Friedmann ones.  But 
calculations indicate that models using smooth 
vs.~totally clumpy distances deviate by less than 20\% in $\vavg$, so 
we will only present results for the smooth distances. 
In general we find that a flat or open clumpy 
model acts intermediate between the smooth cases $\om=0$ and $\om=1$. 

Given the source functions $\Phi(L)$ and $n(z)$ and the cosmological
model, one can determine the distribution and moments of $v$ for
comparison with data:
$$\eqalign{\int_{v_1}^{v_2}dv\,p(v)&= N_{tot}^{-1}\int_0^\infty
dL\,\Phi(L) \int_{z(L,v_1)}^{z(L,v_2)} dz\,(dr_{pr}/dz)
r_a^2(1+z)^3n(z),\cr N_{tot}&=\int_0^\infty dL\,\Phi(L) 
\int_0^{z(L)} dz\,(dr_{pr}/dz)r_a^2(1+z)^3n(z),\cr \langle v\rangle&
=N_{tot}^{-1}\int_0^\infty dL\,\Phi(L)\,D^{-3}(L)\int_0^{z(L)}
dz\,(dr_{pr}/dz) r_a^2r_l^3(1+z)^3n(z).\cr}\eqno(7)$$ 
Here $z(L,v)$ is the inversion of $r_l(z)=D(L)v^{1/3}$ with $z(L)=z(L,1)$ 
and $D$ following from (1) with $C=C_{lim}$. 
The BATSE detector team of the Compton Gamma 
Ray Observatory reports $\langle v\rangle=0.33\pm0.01$. 

[Note that in contrast to the quasar $V/V_{max}$ 
test (Schmidt 1968), uniformity of source spatial distribution 
does not imply $\langle v\rangle=1/2$ and vice versa. 
To see this we write the $z$ integral of (7) in general functional
terms and apply a theorem from integral calculus. 
$$\langle v\rangle=\int_0^Z dz\,f(z)v(z)\bigg/\int_0^Z
dz\,f(z);\eqno(8)$$ 
in (7) we have $f(z)=(n/N_{tot})dV/dz$.  Now (8) has the form of a 
weighted average of $v(z)$ where $v$ takes the values 0 and 1 at the limits. 
However, the theorem states that if the average of a function with that
behavior equals $1/2$ over all intervals, i.e. for all values $Z$,
then the differential of the function must equal the weighting 
function, $f=dv/dz$.  Using $v=(r_l/D)^3$ and equation 
(6) we see this is not generally true for $n$ uniform, so for gamma ray 
bursts the properties of $\langle v\rangle=1/2$ and uniformity are not 
equivalent. 
As a specific example, take the model $\Omega\ll1$, $n(z)=n_0(1+z)^4$: 
one obtains 
$\langle v\rangle=1/2$ for all $z$, since here coincidentally $f=dv/dz$. 
In the quasar test, however, $v$ is defined as $\int_0^V
dV/\int_0^{V_{max}} dV$, so $dv/dz=(\int_0^{V_{max}}dV)^{-1}dV/dz$,
which does equal $f$ for $n$ uniform.] 

Quasilocal cosmological models, those with detection depth $z\ll1$, 
can be severely constrained by calculating 
the volume relation to first order in redshift: 
$$\vavg=(1/2)\bigl[1-(3/7)(1-\beta/4)z\bigr].\eqno(9)$$ 
Here $\beta$ is the density evolution index in 
$n(z)=n_0(1+z)^\beta$, i.e. $\beta=0$ corresponds to no evolution, 
and this first order result is independent of $\om$.  Obtaining 
$\vavg=0.33$ would require $z=0.79$ or negative evolution, the
first violating our low redshift assumption and the second 
requiring excessive evolution at low redshifts (e.g. $\beta=-28$ at 
$z=0.1$ in order to give the value 0.33). 

Were there no problems with threshold bias one could fit simple 
cosmological models with greater depth to the data, obtaining 
the values 
$\vavg=(0.4,0.35,0.3)$ at detection limits $z=(0.7,1.3,2.2)$ for $\om=1$ 
and at $z=(0.7,1.5,3.3)$ for $\om=0$, with no evolution.  Excellent 
approximations to the cosmological model source counts of (7) over 
the redshift range of interest, including evolution, are given to 
$5\%$ by $\langle 
v\rangle=0.5y^{-b}$ where $y=1+z$ and $b=0.36-0.12\beta$ for $\Omega=0$ 
and $b=0.43-0.15\beta$ for $\Omega=1$.  Not only the mean but the binned 
distribution (first line of equation 7) is fit admirably. 
\bigskip 
\leftline{\bf 3. Cosmological parametrization}
\medskip\noindent 
However, two problems exist with using solely the $\vm$ distribution 
to describe the source counts.  One, the information 
from high counts, including the break away from the $-5/2$ slope, is 
scrunched into a small region due to transformation to the $(\cc)^
{-3/2}$ variable.  Two, the threshold variation problem 
of $C_{lim}$ also warps the fundamental $N(C)$ distribution, as clearly 
shown by Petrosian's (1993) deconvolution. 

The definition of the differential source counts is 
$$\eqalign{N(C)&=\int dV\,n(V)\int dL\,\Phi(L)\,\delta(C-L/4\pi r_l^2) 
\cr &=\int dV\,4\pi r_l^2\,n(V)\,\Phi(4\pi r_l^2C).\cr}\eqno(10)$$ 
In the case of uniformity 
(i.e. sources distributed homogeneously in Euclidean space) 
the volume element $dV=4\pi r^2dr$ and $r_l=r$ so 
the integral can be transformed to the form 
$$N(C)=(4\pi)^{-1/2}nC^{-5/2}\int_0^\infty dx\,x^4\Phi(x^2).\eqno(11)$$ 
Although the bright end slope is matched, there is no possibility of a 
break in the power law with respect to $C$, regardless of 
luminosity function.  So a luminosity 
function alone cannot give the observed behavior of the bright end slope of 
$-5/2$ turning over to a shallower faint end slope.  
There are only four possibilities for the cause of a change in slope. 
Changes can occur by varying the geometry, i.e.~$dV$, by  
evolution, i.e.~$n$ or $L(V)$, and by combinations of these with 
each other or with the luminosity function $\Phi(L)$.  

Initially neglect the luminosity function; assume all sources have 
identical luminosity $L$.  Including the appropriate delta function 
for $\Phi(L)$ in (10) gives 
$$N(C)=2\pi n_0 C^{-1}\,[y^{\beta-1}r_l^3(dr_{pr}/dr_l)]_Y,\eqno(12)$$ 
with the bracketed term evaluated at $y=Y$ 
implicitly defined by $L=4\pi Cr_l^2(Y)$. 
The faint slope index becomes constant in the asymptotic, i.e.~high 
redshift, regime with (negative) values found to be $(6+\beta)/4$ or 
$(3+2\beta)/4$ for $\om=0$ or 1. 

The effect of cosmology is that 
as more distant sources are considered the slope of the $\lnc$ 
relation becomes shallower (greater than $-5/2$), i.e. there are fewer 
faint sources than expected because of the limited volume available: 
because of expansion the younger universe was smaller.  Positive density 
evolution (more sources at higher redshift: $\beta>0$), however, 
acts to counteract this.  At the end of this section we examine 
how unreasonable the 
density run has to be to match the source counts. 
Positive luminosity evolution increases $Y$ 
and steepens (shallows) the slope for $\Om=0$ (1).  The effects are 
opposite because the 
two models' volume-redshift dependences are very different. 

If the BATSE data are accepted at face value (i.e.~without threshold 
deconvolution), the faint end slope is 
found to be in the range $-0.8$ to $-1.0$.  In this case low density 
cosmological models are poor fits since substantial 
evolution is needed to achieve such a shallow slope (by redshifts of 
a few rather than asymptotically).  Were this 
extreme evolution in fact to exist, however, the source counts would not 
resemble a broken power law but a {\it very} gradual turnover. 
Constraints such as $\vavg$ from \S2.2 add to the difficulty, even for 
$\om=1$ models. 

To put these constraints on a firmer footing we consider three 
parametrizations of the cosmological $\lnc$ distribution for comparison 
with the data.  Two are the slope 
of the differential counts and the logarithmic range of counts between 
given values of the slope.  The virtue of using slopes and logarithmic 
ranges lies in the ability to neglect questions of absolute values for 
the number of bursts or the peak count rate, made difficult by detector 
dependent effects.  (Of course once a model achieves a fit under these 
conditions then the normalization is important in recovering the 
physical values of source luminosity and density).  
We retain $\vavg$ as a third parameter, 
both for historical reasons and as a complete, rather than point, measure 
of the distribution. 
Although these are not independent, e.g.~knowledge of the slope 
at every point allows reconstruction of the 
(unnormalized) distribution and hence any of the other measures, there 
is sufficient difference in the areas of emphasis each places on the 
counts that it is useful to consider all three. 

We elaborate the source model to include a version of the 
cosmological k-correction, taking into account the finite energy window 
through which the bursts are observed and the redshift of the source 
spectrum.  The relation between the peak count rate $C$ and the 
received flux $f_\nu$ is 
$$C=\int d\nu\,S(\nu)\,(f_\nu/h\nu),\eqno(13)$$ 
where $S(\nu)$ is the window frequency response and $h$ is Planck's 
constant.  The k-correction occurs in transforming the specific 
luminosity $L_\nu$ to $f_\nu$, 
$$f_\nu=(1+z){L[\nu(1+z)]\over4\pi r_l^2}{d[\nu(1+z)]\over d\nu},
\eqno(14)$$ 
where the initial factor of $1+z$ 
accounts for the time dilation since $C$ measures a count rate not number. 
Adopting the spectral models $L_\nu=L_0(\nu/\nu_0)^{-\alpha}$ 
and $S$ flat between $\nu_-$ and $\nu_+$ yields 
$$C=\Bigr[L_0\nu_0^\alpha\int_{\nu_-}^{\nu_+}d\nu\,\nu^{-1-\alpha}/4\pi 
h(\nu_+-\nu_-)\Bigl](1+z)^{2-\alpha}r_l^{-2}(z),\eqno(15)$$ 
generalizing (1). This 
differs from the usual inverse square distance law by the $1+z$ factors; 
one comes from bandwidth dilation, one from time dilation, and the 
$\alpha$ dependence from the spectral shift through the observing window. 
The advantage of using differential logarithmic parameters is that 
the entire group of constants in brackets can be neglected.  

Suppose all sources are 
identical, i.e.~same $L_0$ and $\alpha$, but with evolution possible 
in the source population, as powers of $1+z$ with $\beta$ the index 
for density 
evolution and $\gamma$ for luminosity ($L_0$).  Generalizing (12), 
$$N(C)=4\pi n_0C^{-1}Y^{\beta-1+\delta}(dr_{pr}/dy)\,|d(y^\delta 
r_l^{-2})/dy|^{-1}\equiv AC^{-1}f(Y),\eqno(16)$$ 
where $\delta=\gamma+2-\alpha$ and derivatives are evaluated at $y=Y$, 
the solution to a modified equation (1) [or (15)]: $C=\kappa Y^\delta 
r_l^{-2}(Y)$.  

Taking advantage of the form of (16) in solving for the wanted 
parameters, the (negative) slope of the cumulative distribution $N(>C)$ is 
$$s_c\equiv -{d\ln N(>C)\over d\ln C}={C\,N(C)\over N(>C)}=f(Y)\Big/
\int dr_{pr}y^{-1}r_l^2.\eqno(17)$$ 
This can be seen by direct evaluation of $\int dC\,C^{-1} f(Y)$ or by 
using that $N(>C)$ is just the effective ``volume'' out to $Y(C)$ to 
obtain the denominator.  The differential slope is 
$$s_d\equiv -{d\ln N(C)\over d\ln C}=1-{C\over f}{df\over dy}{dy\over dC}
.\eqno(18)$$ 
Note that because of the bending of the source counts caused by cosmology 
the relation $s_c=s_d-1$ does not generally hold, as it would in the 
scale free case. 

The volume fraction parameter takes its usual definition in terms of counts 
relative to threshold, $\vavg=\langle (\cc)^{-3/2}\rangle$, but note 
that it is now even further from interpretation as a physical volume 
than before. 
Not only is $(\cc)^{-3/2}$ not the volume element (6) 
relative to maximum detectable 
volume (as it would be in the Euclidean case), but it is not even the 
previous distance ratio $(r_l/D)^3$ 
because of the k-correction and time dilation factors.  

Finally, define a count range parameter $R$ by 
$$R(s_d)=\log[C(s_d=2.25)/C(s_d)].\eqno(19)$$ 
This statistic measures how quickly in count space the geometric and 
evolutionary effects become appreciable.  It is interpreted as the range 
{\it to} the highest count rate $C$ where $N(C)$ has slope $s_d$ {\it from} 
where the slope begins to deviate significantly 
($10\%$) from the Euclidean value of 5/2.  

By defining the survey depth 
for each variable -- $z_v$, $z_s$, and $z_r$ -- where the cosmological 
model behavior fits the observations, we discover as viable those 
models which agree on this maxiumum redshift within the data 
uncertainties.  Figure 1 plots the model results for $\om=1$ and 
a variety of evolutionary indices. 

Using data deconvolved 
of threshold effects by the method of Petrosian (1993) suggests 
parameter values in the 
ranges of $s_d=1.6-1.8$, $R=1.0-1.5$, and (independent 
of deconvolution) $\vavg=0.33\pm0.01$ .  None of the models in Figure 1 
simultaneously satisfy these values within the errors, as clearly 
quantified by the depth criterion mentioned above.  
For example, for the $(\beta,\delta)=(-1,0)$ model to achieve 
$\vavg=0.33\pm0.01$ requires $z_v=1.01\pm0.1$ while $s_d=1.7\pm0.1$ 
gives $z_s=0.58\pm0.13$ and $R=1.25\pm0.25$ needs $z_r=0.49\pm0.13$. 
While the plot is for $\Omega=1$, lower densities fit even more poorly. 

Recall that the curves 
incorporate evolution in number density and luminosity, as well as 
k-correction and time dilation, through the parameters $\beta$ and 
$\delta$.  To obtain a unique survey redshift, the curves for 
$\vavg$ and $s_d$ need to be shifted to smaller redshifts, 
which can be accomplished by decreasing $\beta$ or $\delta$; each affects 
the results in roughly the same magnitude.  A unique depth, within the 
data uncertainties, is not achieved until $\beta+\delta<-10$, at which 
point the depth has decreased to $z\approx0.1$.  That requires an 
extreme amount of evolution in a very small redshift range. 

The count range parameter proves very useful, giving a measure of the 
rate of curvature of the count distribution.  Without paying heed to 
this constraint simple, even 
nonevolutionary, models will seem to fit for depths varying from 
$z\approx1-7$.  However the logarithmic count ranges are then 
some two times larger than the values indicated by the deconvolved 
data, i.e.~a factor of 
60-250 off the observed count rate. 

Of course one can always adjust the source spatial distribution 
in (10) so as to obtain any count behavior desired.  
Figure 2 illustrates 
the spatial density $n(z)$ of gamma ray burst sources required to match 
the observed count distribution.  We adopt two forms for this distribution: 
a broken power law with bright slope -2.5 and faint slope -1.7, and a 
more gradual behavior that interpolates between these two 
asymptotes.  For the cosmology we take models with $\Omega=0$ and 
$\Omega=1$ and incorporate 
luminosity evolution, k-correction, and time dilation effects through 
the parameter $\delta$ as before.  The 
requisite cosmological density behavior for explanation of the 
observations 
is not satisfied by the form $n(z)\sim(1+z)^\beta$, which would be a 
straight line on the plot with slope $\beta$, or by any familiar 
cosmological source distribution. 
\bigskip 
\leftline{\bf 4. Conclusion}
\medskip\noindent 
The gamma ray burst source count plot is similar to the classical Hubble 
diagram in that (for the cosmological scenario) the curvature away from 
the Euclidean behavior is a direct measure of cosmological effects.  
The bright end slope of $-5/2$ has a 
natural interpretation in terms of nearby sources, $z<0.1$, where 
geometry and evolution have not yet broken source uniformity.  The faint 
slope value determines a characteristic survey depth, or redshift, as 
does the weighted average $\vavg$ of 
the entire data set, and the observed range in counts between the 
transition region and the asymptotic behavior. 

These three distinct estimates for the depth out to which 
gamma ray bursts are detected -- $z_{slope}$, $z_v$, and $z_{range}$ -- 
must be equal for the cosmological model fit.  
Figure 1 shows, however, that this does not hold for any of the models 
considered with power law evolution.  Figure 2 illustrates the run of 
spatial number density required for 
cosmological sources to match the observed count distribution, again 
in disagreement with power law evolution. 
The models took into account the cosmological characteristics of geometry, 
k-correction and time dilation, as well as 
number and luminosity evolution power law in redshift. 

\bigskip 
\noindent{\it Acknowledgments.} 
I am indebted to Dieter Hartmann, Wlodek Kluzniak, and Vah\'e Petrosian for 
helpful discussions.  Use was made of the BATSE online database at 
gronews@ grossc.gsfc.nasa.gov. 
\bigskip  
\leftline{\bf References} 
\smallskip 
\parindent=0in 
Caditz, D., Petrosian, V., 1993, ApJ, 416, 450\par 
Dyer, C.C., Roeder, R.C., 1973, ApJ, 180, L31\par 
Ehlers, J., Schneider, P., 1986,  A\&A, 168, 57\par 
Petrosian, V., 1993, ApJ, 402, L33\par 
Schmidt, M., 1968, ApJ, 151, 393\par 
Schmidt, M., Higdon, J.C., Hueter, G., 1988, ApJ, 329, L85\par 
Schneider, P., 1987, ApJ, 316, L7\par 
\medskip 
\leftline{\bf Figure captions} 
\smallskip\noindent 
{\bf Fig. 1.} The statistical parameters, normalized to the center 
of their observed range, are plotted vs. redshift.  Those three curves 
starting at $z=0$ with value 1.52 show $\vavg$, those with value 1.47 
show $s_d$, and the positive slope curves show $R$.  Solid curves have 
($\beta,\delta$)=(0,0), long dashed have (0,-1), and short dashed 
(-1,0).  The intersection of each curve with the value one defines the 
appropriate depth $z_v$, $z_s$, or $z_r$.  A successful model would have 
all solid curves, say, intersecting the dotted line simultaneously. 
Of course uncertainty in the data broadens the dotted line, into 
$1\pm0.03$ for $\vavg$, $1\pm0.06$ for $s_d$, and $1\pm0.2$ for $R$. 
\smallskip\noindent 
{\bf Fig. 2.} The cosmological spatial number density derived to 
match a source counts model is plotted vs.~redshift.  
The solid and dotted lines use 
a broken $N(C)$ and $\Om=0$ and 1, respectively; the short and long 
dashed lines 
a gradual $N(C)$ and $\Om=0$ and 1, respectively. 
The break, or turnover, in counts is assumed to lie 
at $z=0.1$; it is actually a function of the intrinsic luminosity and 
spectrum and will scale the horizontal axis accordingly, but should lie 
close to the value adopted. 
\vfill\break
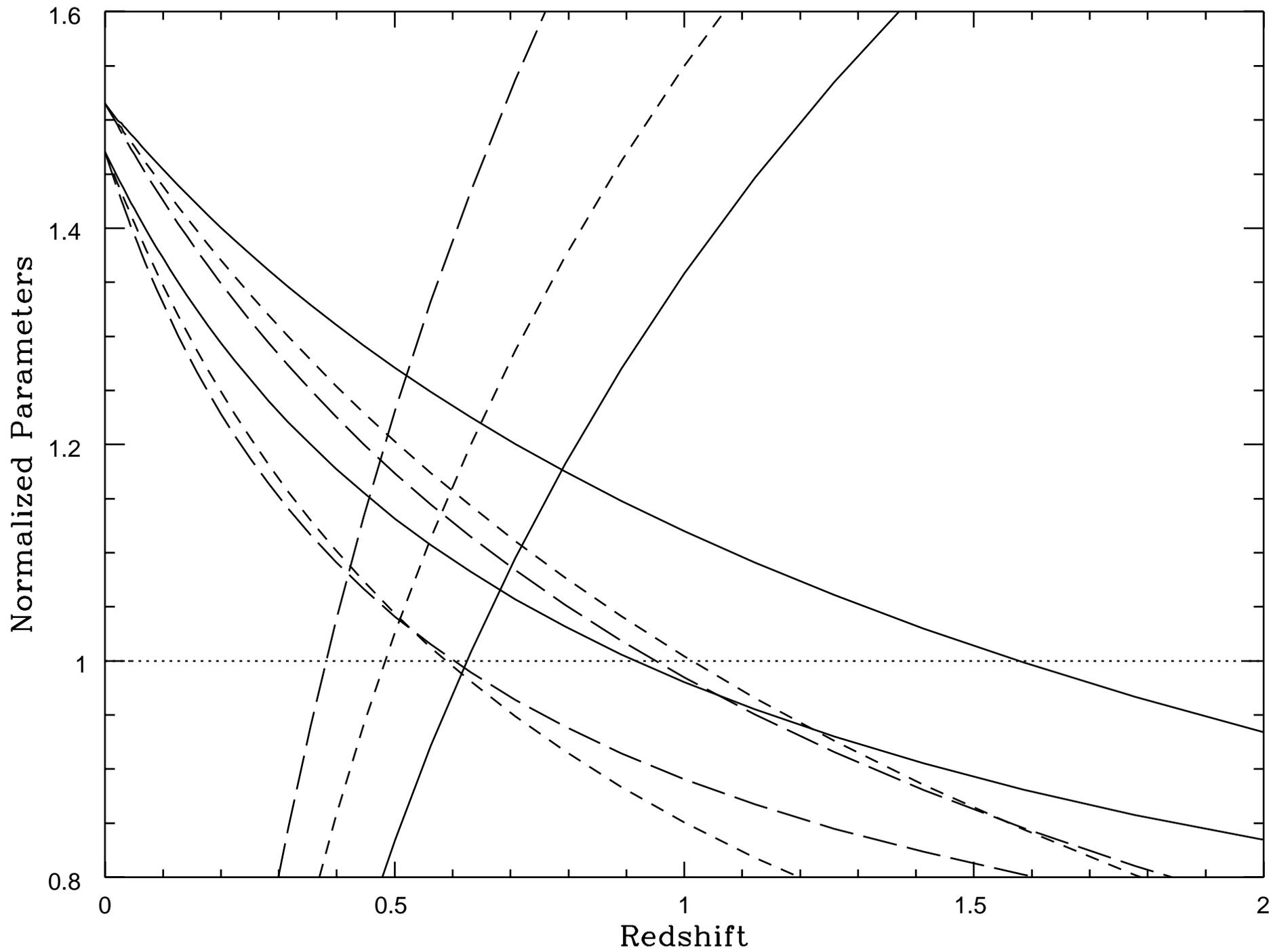
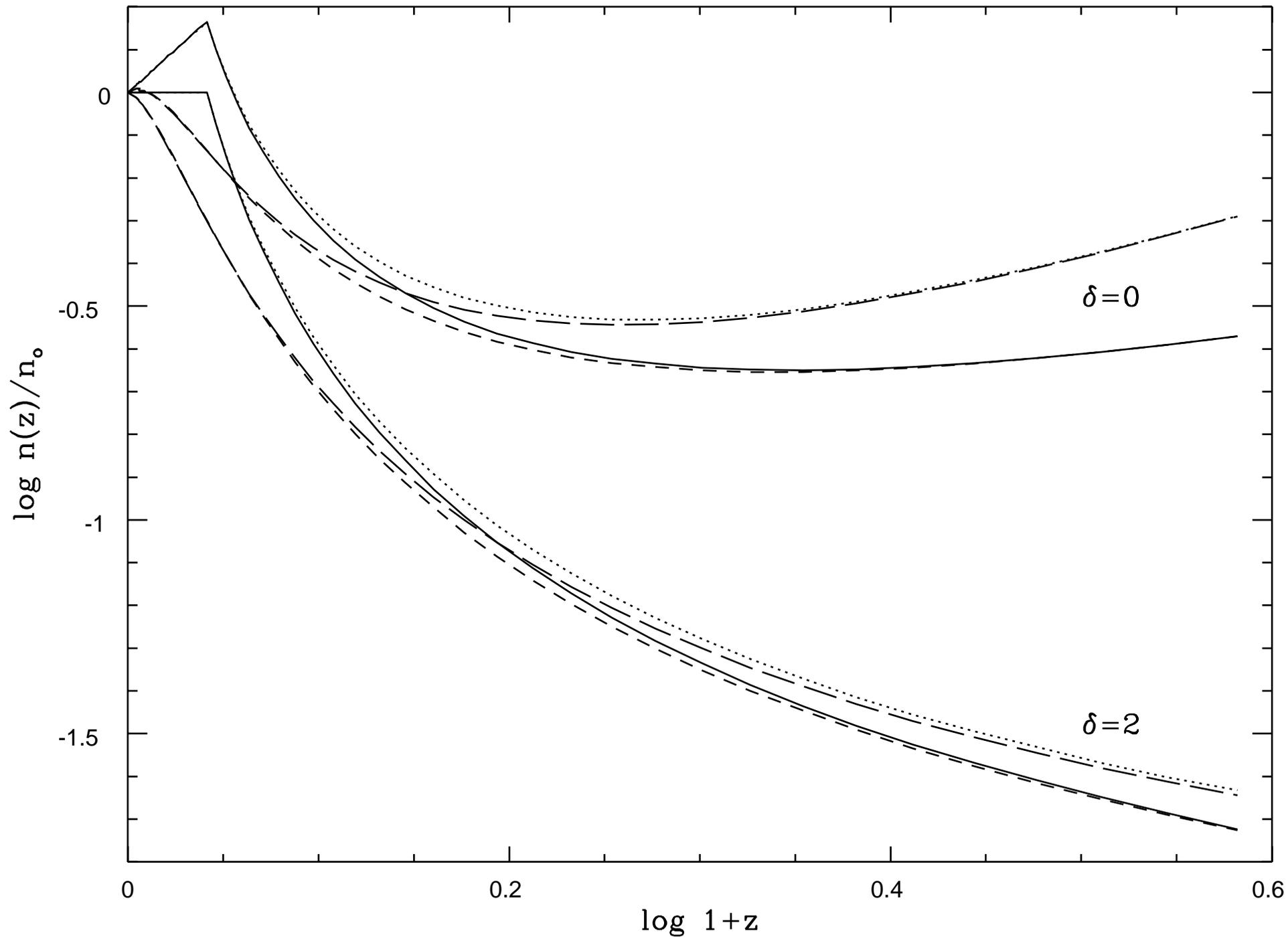
\bye